# Entanglement of *N* distinguishable particles

Tomasz Bigaj

Warsaw University

**Abstract**. In their 2002 article, Ghirardi, Marinatto and Weber have proposed a formal analysis of the entanglement properties for a system consisting of *N* distinguishable particles. Their analysis leads to the differentiation of three possible situations that can arise in such systems: complete entanglement, complete non-entanglement, and the remaining cases. I argue that this categorization leaves out one important possibility in which a system is completely entangled, and yet some of its subsystems are mutually non-entangled. As an example I present and discuss a state of a three-particle system which cannot be decomposed into two non-entangled systems, and yet particle number one is not entangled with particle number three. Consequently, I introduce a new notion of utter entanglement, and I argue that some systems may be completely but not utterly entangled.

The notion of entanglement remains at the centre of the foundational analysis of quantum mechanics. To date, one of the most comprehensive studies of mathematical and conceptual features of quantum entanglement in various settings is the 2002 paper co-authored by G. Ghirardi, L. Marinatto and T. Weber [1]. One section of this extensive article has been devoted to the analysis of the entanglement relations that can occur in a system containing *N* distinguishable particles. Because *N* particles can remain in different entanglement settings relative to one another, we need to distinguish various types of entanglement relations that may emerge in the entire composite system. Ghirardi, Marinatto and Weber (henceforth referred to as GMW) formulate precise mathematical definitions of such possible categories of entanglement, including cases of complete entanglement and complete non-entanglement. However, it turns out that their categorization is not exhaustive. This article contains an attempt to amend GMW's analysis in order to make it more comprehensive.

In the first section I briefly outline the original method of analysing possible correlations among *N* particles proposed by GMW. The second section sketches a proof, missing from GMW's article, that their procedure is consistent. In the third section I present a case of a three-particle system prepared in a state such that although the system as a whole cannot be bipartitioned into two non-entangled subsystems (and hence qualifies as completely entangled), two particles within the system are arguably not entangled with one another. An interesting physical realisation of such a situation is provided by interpreting the states of the particles as consisting of spatial and internal (spin) degrees of freedom. In that case the mathematical form of the initial state implies that particles 1 and 2 have their spins entangled, while the entanglement of particles 2 and 3 affects only their positions. In the fourth section I argue that this new case cannot be classified with the help of another distinction introduced by GMW between partially and totally entangled systems. To categorize it, I introduce a new notion of utter entanglement, showing that complete entanglement does not have to be utter.

**1.** Following GMW, our main goal will be to categorize all possible entanglement relations that may arise in a composite system consisting of $N$ distinguishable particles. The starting assumption is that the system $S$ is prepared in a pure state described by the vector $|\psi(1,...,N)\rangle$. This state, in turn, determines the states of all subsystems of $S$, which are obtained by reducing $|\psi(1,...,N)\rangle$. The general method of getting the reduced states is by applying the partial trace operation to $|\psi(1,...,N)\rangle$. Thus, the subsystem $S_{(1...M)}$ consisting of particles 1, 2, ..., $M$, where $M < N$, will be assigned the state represented by the following density operator

$$\rho^{(1...M)} = Tr^{(M+1...N)}(|\psi(1,...,N)\rangle\langle\psi(1,...,N)|)$$

where $Tr^{(M+1...N)}$ is the partial trace calculated over the spaces corresponding to the particles $M+1$, ..., $N$. It is worth noting that the state assigned to a given subsystem $S_{(1...M)}$ is independent from what system $S_{(1...M)}$ is considered to be a subsystem of. That is, if we decide first to calculate, using the above formula, the reduced density operator for a bigger subsystem consisting of particles 1, ..., $M$, $M+1$, ..., $K$, and then we apply the same procedure to reduce the resulting state to the subsystem $S_{(1...M)}$, the final state will be precisely the same as above. This follows directly from the fact that the application of two partial trace operations is equivalent to one partial trace operation over the sum of both systems associated with the separate trace operations.

The first question we have to ask with respect to the global entanglement of *S* is whether it is possible to decompose it into two subsystems such that they are not entangled with each other. There are several equivalent ways of presenting the condition of non-entanglement between two subsystems containing particles 1, ..., *N* and *N*+1, ..., *M*. The most popular definition of non-entanglement is based on the factorizability condition.

Def. 1. The subsystem $S_{(1...M)}$ is non-entangled with the subsystem $S_{(M+1...N)}$ iff there exist vectors $|\lambda(1,...,M)\rangle$ and $|\varphi(M+1,...,N)\rangle$, representing possible states of $S_{(1...M)}$ and $S_{(M+1...N)}$ respectively, such that $|\psi(1,...,N)\rangle = |\lambda(1,...,M)\rangle \otimes |\varphi(M+1,...,N)\rangle$.

Another possible definition of non-entanglement uses the notion of the reduced state.

Def. 2. The subsystem $S_{(1...M)}$ is non-entangled with the subsystem $S_{(M+1...N)}$ iff the reduced density operator $\rho^{(1...M)}$ is a projection operator onto a one-dimensional subspace of the space $H_1 \otimes H_2 \otimes ... \otimes H_M$ ($\rho^{(1...M)}$ can be presented as $|\lambda(1,...,M)\rangle\langle\lambda(1,...,M)|$)

Other equivalent definitions of non-entanglement are possible too, but we won't write them down, referring the reader to the literature instead.[1]

The procedure used by GMW in order to analyze the entanglement of the composite system *S* consisting of *N* particles is quite straightforward. First, we have to check whether it is possible to split *S* into two non-entangled subsystems *S′* and *S″*. If this can be done, then the procedure has to be repeated for each of the subsystems *S′* and *S″* in order to bipartition them into even smaller subsystems not entangled with one another, if possible. That way we can arrive at the finest partitioning of *S* into several independent subsystems $S_1, S_2, ..., S_k$ such that none of the subsystems $S_i$ is further decomposable into non-entangled components. Now two possibilities have to be considered. One is that the systems $S_1, S_2, ..., S_k$ may turn out to be one-particle systems. This means that the initial system *S* is *completely unentangled*, and each particle constituting it has its own pure state. In other words, the state vector $|\psi(1,...,N)\rangle$ can be presented as the product of *N* vectors each belonging to a one-particle Hilbert space. But it is also possible that *S* does not have any non-entangled subsystems, i.e.

---

[1] The main definition of non-entanglement given in [1, p. 68] refers to the existence of a one-dimensional projection operator characterising the subsystem $S_{(1...M)}$, whose expectation value in the initial state is 1. Definitions of non-entanglement based on the notion of the Schmidt number and von Neumann entropy are mentioned in [2, p. 012109-4]. Another popular criterion of non-entanglement is that the trace operator of the square of the reduced density operator should equal one (cf [3, p. 50]).

there is only one system in the set of non-decomposable subsystems $S_1, S_2, ..., S_k$, and this system is $S$ itself. In this case $S$ is said to be *completely entangled*.

This distinction can be conveniently presented as follows. In accordance with the adopted notation let $k$ be the number of mutually non-entangled subsystems of $S$ which are not decomposable into further non-entangled parts. Then, if $k = N$, the system $S$ is completely non-entangled, and if $k = 1$, $S$ is completely entangled. If $k$ falls between 1 and $N$, we have a case in which $S$ is decomposable into non-entangled composite subsystems which themselves are completely entangled.

**2.** It turns out, however, that the above analysis has to be amended in two respects. First, let us start with a relatively minor issue. We have to make sure that the procedure of identifying the smallest entangled components of a given system is consistent, i.e. that it leads to the unique outcome which is independent of the initial separation into two non-entangled subsystems. The uniqueness property can be argued for as follows. Suppose that it is possible to make two bipartitions of $S$ into subsystems $S_K$ and $S_{K'}$ and into subsystems $S_L$ and $S_{L'}$, and that both pairs $S_K, S_{K'}$ and $S_L, S_{L'}$ are mutually non-entangled. To ensure the uniqueness of the procedure of separation into smallest non-entangled components of $S$, we have to prove that the subsystems $S_{KL} = S_K \cap S_L$, $S_{KL'} = S_K \cap S_{L'}$, $S_{K'L} = S_{K'} \cap S_L$, $S_{K'L'} = S_{K'} \cap S_{L'}$ are also mutually non-entangled. That way we can argue that no matter which initial bipartition we start with, we will end up with the same decomposition into the smallest mutually non-entangled subsystems of the system $S$.

A proof of the above-mentioned fact can be sketched as follows. By assumption the initial state of the system $S$ factorizes into the product of the components describing the states of $S_K, S_{K'}$ and $S_L, S_{L'}$ respectively:

$$|\psi(1, ..., N)\rangle = |\psi\rangle_K |\psi\rangle_{K'} = |\psi\rangle_L |\psi\rangle_{L'}$$

Now we can write down the Schmidt decompositions for the vectors $|\psi\rangle_K$ and $|\psi\rangle_{K'}$ in the bases of subsystems $S_{KL}, S_{KL'}$ and $S_{K'L}, S_{K'L'}$.

$$|\psi\rangle_K = \sum_n a_n |\lambda_n\rangle_{KL} |\varphi_n\rangle_{KL'}$$

$$|\psi\rangle_{K'} = \sum_l b_l |\chi_l\rangle_{K'L} |\mu_l\rangle_{K'L'}$$

The state vector of the system $S$ can be thus presented as follows:

$$|\psi(1, ..., N)\rangle = \sum_{nl} a_n b_l |\lambda_n\rangle_{KL} |\varphi_n\rangle_{KL'} |\chi_l\rangle_{K'L} |\mu_l\rangle_{K'L'}$$

But we know that $|\psi(1,...,N)\rangle$ factorizes into the direct product of vectors $|\psi\rangle_L$ and $|\psi\rangle_{L'}$. This is possible only when all coefficients $a_n$ and $b_l$ but one equal zero. But in this case clearly $|\psi(1,...,N)\rangle$ decomposes into the product of four vectors, describing the states of the subsystems $S_{KL}$, $S_{KL'}$, $S_{K'L}$, and $S_{K'L'}$. Therefore these subsystems are not entangled.

**3.** However, the analysis proposed by GMW is incomplete in a more fundamental way. Their classification into completely entangled and completely unentangled systems overlooks the fact that even if the system $S$ is not fully decomposable into two non-entangled subsystems, there may be some 'pockets' of mutually non-entangled subsystems within $S$ left. This is possible, because when a given subsystem $S'$ receives a reduced density operator $\rho'$ as the representation of its state, $\rho'$ may turn out to be the product of two density operators $\rho_1'$ and $\rho_2'$ each representing the state of one subsystem of $S'$. In such a case the subsystems have to be deemed non-entangled (cf. [3, p. 52]). Below I will present and carefully examine an example of such a situation. This example involves three particles whose state spaces are four-dimensional Hilbert spaces spanned by orthonormal vectors $|0\rangle, |1\rangle, |2\rangle, |3\rangle$. The considered state of the system $S$ is given as follows:

(*) $$|\psi(1,2,3)\rangle = \frac{1}{2}(|0\rangle_1|1\rangle_2|2\rangle_3 + |0\rangle_1|3\rangle_2|0\rangle_3 + |1\rangle_1|0\rangle_2|2\rangle_3 + |1\rangle_1|2\rangle_2|0\rangle_3)$$

We can now calculate the reduced density operators for particles 1, 2 and 3 separately.

$$\rho_1 = Tr^{(2,3)}(|\psi(1,23)\rangle\langle\psi(1,2,3)|) = \frac{1}{2}(|0\rangle\langle 0| + |1\rangle\langle 1|)$$

$$\rho_2 = Tr^{(1,3)}(|\psi(1,23)\rangle\langle\psi(1,2,3)|) = \frac{1}{4}(|0\rangle\langle 0| + |1\rangle\langle 1| + |2\rangle\langle 2| + |3\rangle\langle 3|)$$

$$\rho_3 = Tr^{(1,2)}(|\psi(1,23)\rangle\langle\psi(1,2,3)|) = \frac{1}{2}(|0\rangle\langle 0| + |2\rangle\langle 2|)$$

Clearly, all reduced one-particle states are mixed rather than pure, and therefore the system $S$ cannot be decomposed into non-entangled subsystems. However, let us now calculate the reduced density operator for the two-particle subsystem $S_{(1,3)}$:

$$\rho_{1,3} = Tr^{(2)}(|\psi(1,23)\rangle\langle\psi(1,2,3)|) = \frac{1}{4}(|0\rangle_{11}\langle 0|\otimes|2\rangle_{33}\langle 2| + |0\rangle_{11}\langle 0|\otimes|0\rangle_{33}\langle 0| +$$
$$+ |1\rangle_{11}\langle 1|\otimes|2\rangle_{33}\langle 2| + |1\rangle_{11}\langle 1|\otimes|0\rangle_{33}\langle 0|) =$$
$$= \frac{1}{4}(|0\rangle_{11}\langle 0| + |1\rangle_{11}\langle 1|)\otimes(|2\rangle_{33}\langle 2| + |0\rangle_{33}\langle 0|) = \rho_1\otimes\rho_3$$

Because the reduced state $\rho_{1,3}$ is the product of the states of particle 1 and 3, it has to be concluded that 1 is not entangled with 3. Thus we have an interesting case of entanglement here. Particle 1 is entangled with the subsystem containing particles 2 and 3, but this entanglement affects only the relation between 1 and 2, not 1 and 3. In particular, no non-local correlations can be detected between outcomes of measurements performed on particles 1 and 3. Similarly, the entanglement of particle 2 with the two-particle system {1, 3} arises entirely in virtue of the entanglement between 2 and 3. It can be verified by analogous calculations that particle 1 is entangled with 2, and 2 is entangled with 3, as neither reduced density operator $\rho_{1,2}$ nor $\rho_{2,3}$ factorizes. But clearly the relation of entanglement is not transitive, hence 1 and 3 may be, and actually are, non-entangled.

The state $|\psi(1,2,3)\rangle$ can be given a suggestive physical interpretation when we identify the vectors with states having both internal and spatial degrees of freedom. Let us assume that the particles can be characterized by their spin-half values up ($|\uparrow\rangle$) and down ($|\downarrow\rangle$), and by their two possible locations left ($|L\rangle$) and right ($|R\rangle$). In addition, let us make the following identifications:

$$|0\rangle = |R\rangle|\uparrow\rangle$$
$$|1\rangle = |R\rangle|\downarrow\rangle$$
$$|2\rangle = |L\rangle|\uparrow\rangle$$
$$|3\rangle = |L\rangle|\downarrow\rangle$$

Under this interpretation the initial state of the system (∗) can be rewritten in the form of the following vector:

(∗∗) $\quad |\psi(1,2,3)\rangle = |R\rangle_1(|\uparrow\rangle_1|\downarrow\rangle_2 + |\downarrow\rangle_1|\uparrow\rangle_2)(|R\rangle_2|L\rangle_3 + |L\rangle_2|R\rangle_3)|\uparrow\rangle_3$

The mathematical form of the above vector already suggests the interpretation according to which the spins of particles 1 and 2 and positions of particles 2 and 3 are entangled, while particle 1 has a precise location and particle 3 has a precise spin. Calculation of reduced density matrices confirms this observation:

$$\rho_1 = |R\rangle\langle R|(\frac{1}{2}|\uparrow\rangle\langle\uparrow| + \frac{1}{2}|\downarrow\rangle\langle\downarrow|)$$

$$\rho_2 = \frac{1}{4}(|R\rangle\langle R| + |L\rangle\langle L|)(|\uparrow\rangle\langle\uparrow| + |\downarrow\rangle\langle\downarrow|)$$

$$\rho_3 = (\frac{1}{2}|R\rangle\langle R| + \frac{1}{2}|L\rangle\langle L|)|\uparrow\rangle\langle\uparrow|$$

The reduced state for particle 1 is a mixture of spins but its location is precisely *R*, whereas particle 3 has the precise spin up, but its location is a mixture of *R* and *L*. Particle number 2 has neither spin, nor position well-defined. Particle 2 is entangled both with 1 (via spins) and with 3 (via positions). But no direct entanglement between particles 1 and 3 is present. By looking at the formula (∗∗) we can immediately see that a measurement of spin on particle 1 changes non-locally the spin state of particle 2 (forcing it to admit one of the two definite values depending on the outcome), but doesn't affect the state of particle 3. On the other hand, a position measurement performed on particle 3 affects the position of particle 2 without influencing in any way the reduced state of particle 1.

**4.** It may be observed that the entanglement between system $S_1$ and system $S_{(2,3)}$, as well as between $S_3$ and $S_{(1,2)}$, is of the type that GMW call *partial entanglement* (cf. [1, p. 69]). The general definition of partial entanglement is as follows.

Def. 3  The subsystem $S_{(1...M)}$ is partially entangled with the subsystem $S_{(M+1...N)}$ iff the range of the reduced density operator $\rho^{(1...M)}$ is a proper submanifold (whose dimensionality is greater than one) of the total state space $H_1 \otimes H_2 \otimes ... \otimes H_M$.

If Def. 3 is satisfied, the entangled systems can be ascribed some definite properties in the form of projection operators which are projecting onto a subspace which is more than one-dimensional, but does not coincide with the entire state space. In our case the range of the operator $\rho_1$ describing the state of the first particle is a proper subset of the entire state space, as it coincides with the product of the entire spin space and the one-dimensional ray spanned by vector $|R\rangle$. Analogously, the range of $\rho_3$ is the product of the whole two-dimensional position space and the one-dimensional ray spanned by $|\uparrow\rangle$. Consequently, particle 1 is only partially entangled with the remaining subsystem, and so is particle 3. In contrast with this, the density operator $\rho_2$ for particle number 2 has its range identical with the product of two entire spaces for spins and positions. As a result, no definite property can be associated with this system, and in GMW's terminology particle 2 is *totally* (i.e. not partially) entangled with the system consisting of particles 1 and 3.

However, it would be incorrect to claim that the special character of the entanglement of the state (∗∗) can be fully expressed by categorizing it as a case of complete but not total entanglement. It can be easily verified that there are completely and not totally entangled states which nevertheless lack the unique feature of the state (∗∗), i.e. the non-entanglement of some small subsystems within the entire completely entangled system. Consider, for instance, the following three-particle state:

$$|\psi(1,2,3)\rangle = (|\uparrow\rangle_1|\uparrow\rangle_2|\uparrow\rangle_3 + |\downarrow\rangle_1|\downarrow\rangle_2|\downarrow\rangle_3)|A\rangle_1|B\rangle_2|C\rangle_3$$

where *A*, *B*, *C* denote three distinct locations. It is clear that the three particles are not totally entangled, as their positions are well-defined, and yet each particle is entangled with any other particle (the spin measurement on any particle changes the state of the remaining two). In order to distinguish this case from the cases similar to (∗∗), we should introduce a new category of entanglement – let's call it *utter entanglement* – with the help of the following definition.

Def. 4. A composite system *S* is utterly entangled iff *S* is completely entangled and for every proper subsystem *S′* of *S*, its state $\rho'$ cannot be written in the form $\rho_a \otimes \rho_b$, where $\rho_a$ and $\rho_b$ are states of the subsystems composing *S′*.

As we know from the above-mentioned example, there are states which are completely but not utterly entangled. The impossibility of dividing a system *S* into two non-entangled subsystems does not imply that every subsystem of *S* is entangled with every other subsystem.

In conclusion, we can distinguish the following categories of entanglement which can occur in a system consisting of *N* distinguishable particles. To begin with, the system can be completely unentangled, which means that its state is a direct product of *N* states of separate particles. If the system can be split into *k* subsystems ($k < N$) which are mutually non-entangled but cannot be further divided into non-entangled components, this is a case of incomplete entanglement. A system which cannot be divided into two non-entangled subsystems is called completely entangled. Within the category of completely entangled systems we can distinguish systems which are not utterly entangled, i.e. such that they still contain two or more subsystems (which however do not jointly compose the entire system) which are not entangled with one another. The last category of entanglement is utter entanglement, which means that every subsystem is entangled with every other subsystem. Finally, it should be added that the concept of total entanglement as presented above is orthogonal to the introduced distinctions. That is, each of the above-mentioned cases of entanglement can be a case of total or partial entanglement.